\documentclass[twocolumn,superscriptaddress,showpacs,aps,prb]{revtex4}
\usepackage{makeidx}
\usepackage{bm}
\usepackage{graphics}
\usepackage[dvips]{epsfig}
\newcounter{fig}

\begin{document}
\title{Symmetry constraints on phonon dispersion in graphene  }
\author{L.A. Falkovsky}
\affiliation{L.D. Landau Institute for Theoretical Physics, Moscow
117334, Russia} \affiliation{Institute of the High Pressure
Physics, Troitsk 142190, Russia}

\begin{abstract}
Taking into account the constraints imposed by the lattice
symmetry, we  calculate  the phonon dispersion  for graphene with
interactions between the first, second, and third nearest
neighbors in the framework of the Born--von Karman model.
Analytical expressions obtained for the dispersion of the
out-of-plane (bending) modes give the nonzero sound velocity. The
dispersion of  four in-plane modes is determined by coupled
equations.  Values of the force constants are found in fitting
with frequencies at critical points and with elastic constants
measured on graphite.
\end{abstract}
\maketitle

\section{Introduction}
Since the pioneering experiments on   graphene (a single atomic
layer of graphite) \cite{Novo, ZSA}, main attention has been
devoted to its electronic properties. More recently, Raman
spectroscopy \cite{FMS} extends to investigations of the lattice
dynamics of graphene. It was found that the frequency ($\approx$
1590 cm$^{-1}$) of the Raman mode  in graphene agrees with its
value in graphite. Also, the overtone of the $D$ mode visible
almost in all carbon-consisting materials was observed   at about
2600 cm$^{-1}$. However, this information is very  meagre and does
not provide a way to describe the lattice dynamics. The detailed
knowledge of the lattice dynamics  and electron-phonon
interactions \cite{NG} is needed for interpretations of the Raman
scattering as well as of the transport phenomena.

Several models \cite{DL,NWS,AR,NB,MKH,AD,GMR} have been proposed
to predict the phonon dispersion in graphene and bulk graphite
from empirical force-constant calculations.
A simplest approach assumes the diagonal form of the
force-constant matrix which contains   three constants  for the
interaction of an atom with all its $n$th-nearest neighbor. Thus,
we meet  12 constants for graphene in the popular 4th-nearest
neighbor approach or 15 constants in the 5th-nearest neighbor one
\cite{MMD}. The number of constants could be diminished if the
model interactions are used \cite{BO,LDC,MCZ} or if the phonon
dispersion is considered only for the distinctive directions in
the Brillouin zone \cite{OA}.

On the other hand, we can use the most recent results
\cite{DK,WR,MRTR,MMD,MM,PL} of the first-principal calculations
for the phonon dispersion in graphene and graphite.
Comparison of that results for the high-frequency modes (see Table
1) shows  disagreements as large as 50 cm$^{-1}$ between the
various approaches. The discrepancies could come either from an
assumption that  the force-constant matrix for the atom-neighbor
interaction has a diagonal form or from an overestimation of the
low-frequency modes. It is evident that atoms  move more freely in
out-of-pane direction in graphene than in graphite. Therefore, the
frequencies of the out-of-plane mode in graphene should be less
than the corresponding frequencies in graphite. Moreover, if the
stiffness of the graphene layer is neglected, the dispersion of
the acoustic out-of-plane mode becomes quadratic as  seen from the
equation of elasticity (see also Ref. \cite{DS}).
The interaction between layers in graphite can be estimated from
the splitting of the low-frequency ZA and ZO$^{\prime}$ modes in
graphite. One can see,
 for instance, from Ref. \cite{MM} that the value of
 the splitting is
 as much as 130 cm$^{-1}$. It means that (i)  the result of graphene
 stiffness cannot be larger than that interaction and (ii) the
 agreement between the theory for graphene and the experimental
 low-frequency data for graphite cannot be better than about 130 cm$^{-1}$.

 Here we present an analytical description   of the phonon
dispersion in graphene. This is done
 within the framework of  the  Born--von Karman model for
 the honeycomb graphene lattice including interactions only with
  first, second, and third nearest neighbors and taking the
 constraints imposed by the lattice symmetry
  into account.
 We show that the out-of-plane (bending)  and in-plane modes are
  decoupled from each other. The out-of-plane modes are described by
  three force-constants  determined in fitting with
  the  Raman frequency and  smallest
  elastic constant $C_{44}$.
  In the narrow  wave-vector interval near the $\Gamma$ point,
  the acoustic out-of-plane mode has a linear dispersion with the
  nonzero sound velocity. This means that  a single graphene layer possesses
  the small but finite stiffness  in  contradiction with results
  of Ref. \cite{DS}
We should emphasize that the quadratic dispersion of the acoustic
mode leads to the large contribution (proportional to the sample
size squared  of  the long-range fluctuations, that is much
stronger than the logarithmic function  in the case of the linear
dispersion. Six force-constants describing the in-plane modes are
found in fitting with their   frequencies in the critical points
and elastic constants $C_{11}$ and $C_{12}$ of graphite.
   \section{Phonon dynamics in nearest neighbor approximation}
  The equations of motion in the harmonic
approximation are written in the well-known form
\begin{eqnarray} \label{eqmo}
&\sum\limits_{j, m, \kappa'}\Phi^{\kappa \kappa'}_{ij}({\bf a}_n -
{\bf a}_m) u_j^{\kappa'} ({\bf a}_m)=\omega^2 u_i^{\kappa}({\bf
a}_n) ,
\end{eqnarray}
where the vectors ${\bf a}_{n}$ numerate the lattice cells, the
superscripts $\kappa, \kappa'$  note two sublattices $A$ and $B$,
and the subscripts $i, j = x, y, z$ take three values
corresponding to the space coordinates. Since the potential energy
is the quadratic function of the atomic displacements
$u^{A}_{i}({\bf a}_n) $ and $u^{B}_{i}({\bf a}_n) $, the
force-constant matrix can be taken in the symmetric form,
$\Phi_{ij}^{A B}({\bf a}_{n}) = \Phi_{ji}^{B A}(-{\bf a}_{n}),$
and its Fourier transform, i.e. the dynamical matrix, is a
Hermitian matrix.

Each atom, for instance, ${\bf A}_0$ (see Fig. 1) has  three first
neighbors in the other sublattice, {\it i.e.} $B$, with the
relative vectors ${\bf B}_{1}=a(1, 0),\, {\bf
B}_{2,3}=a(-1,\pm\sqrt{3})/2,$ where $a = 1.42 \AA$ is  the
carbon-carbon distance. The  second neighbors are in the same
sublattice $A$ at distances $\sqrt3 a$ with the relative
 vectors
${\bf A}_{1,4}=\pm a(0, \sqrt3),\, {\bf A}_{2,5}=\pm
a(-3,\sqrt3)/2,\,{\bf A}_{3,6}=\mp a(3,\sqrt3)/2.$ The distance
$2a$ to the third neighbors  $ {\bf B}_{1}^{\prime}=a(2,0), \,{\bf
B}_{2,3}^{\prime}=a(1, \mp\sqrt3)$ is slightly larger. The
distance to the fourth neighbors is $\sqrt7 a=2.65 a$. So,  the
difference between distances to  the third and to fourth
 neighbors is nearly the same as the
difference between distances to the first and to second
 ones. We will see that the force-constants become less by factor 5
 while going from the first to the second neighbors (see Table 3).
 Therefore, we do not include the fourth neighbors into consideration.

\begin{figure}[b]\resizebox{.2\textwidth}{!}{\includegraphics{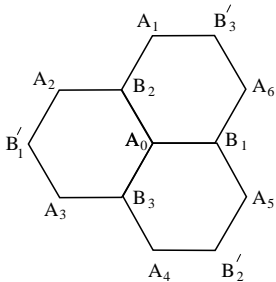}}
\caption{\label{1} First, second, and third neighbors in the
graphene lattice.}
\end{figure}
For the first and third neighbors (in the $B$ sublattice), the
dynamical matrix  has the form
\begin{equation}\label{sc}
 \begin{array}{c}
\phi^{AB}_{i j}({\bf q}) = \sum_{\kappa=1}^3 \Phi^{AB}_{i j}({\bf
B}_{\kappa})\exp(i{\bf qB}_{\kappa})\\
+\sum_{\kappa=1}^3\Phi^{AB}_{i j}({\bf
B}_{\kappa}^{\prime})\exp(i{\bf qB}_{\kappa}^{\prime}),
\end{array}
\end{equation}
and
 for the second neighbors (in the $A$ sublattice)
\begin{eqnarray} \label{ss}
&\phi^{AA}_{i j}({\bf q})= \Phi^{AA}_{i j}({\bf
A}_{0})+\sum_{\kappa=1}^6 \Phi^{AA}_{i j}({\bf
A}_{\kappa})\exp(i{\bf qA}_{\kappa}),
\end{eqnarray}
where ${\bf A}_{0}$ indices the atom chosen at the center of the
coordinate system in the $A$ sublattice and the wave vector ${\bf
q}$ is taken in units of $1/a$.

The point  group $D_{6h}$ of the honeycomb lattice is generated by
$\{C_6,\sigma _v,\sigma _z\}$, where   $\sigma_ z$ is a reflection
$z\to -z$ by the plane that contains the graphene layer, $C_6$ is
a rotation by $\pi/3$ around the $z$ axis, and $\sigma_v$ is a
reflection by the $xz$ plane. The transformations of the group
impose constraints on the dynamical matrix. To obtain them, we
introduce variables $\xi,\eta=x\pm iy$ transforming under the
rotation $C_3$ around the $z$-axis
 (taken at the ${\bf A}_{0}$ atom) as follows
 $(\xi,\eta)\rightarrow (\xi,\eta)\exp(\pm 2\pi i/3)$. In
the  rotation, the atoms change their positions ${\bf
B}_1\rightarrow {\bf B}_2\rightarrow{\bf B}_3$,  ${\bf
A}_1\rightarrow {\bf A}_3\rightarrow{\bf A}_5$, and ${\bf
A}_2\rightarrow {\bf A}_4\rightarrow{\bf A}_6$. Therefore, all the
force constants $\Phi^{AB}_{\xi \eta}({\bf B}_{\kappa})$ with the
different $\kappa$ (as well as $\Phi^{AB}_{z z}({\bf
B}_{\kappa})$) are equal  to one another, but the force constants
with the coincident  subscripts $\xi$ or $\eta$  transform as
covariant variables. For instance,
$\Phi^{AB}_{\xi\xi}({\bf B}_{1})=\Phi^{AB}_{\xi\xi}({\bf
B}_{2})\exp{(2\pi i/3)}
=\Phi^{AB}_{\xi\xi}({\bf B}_{3})\exp{(-2\pi i/3)}.$

 The relation between
$\Phi^{AA}_{\xi\xi}({\bf A}_{\kappa})$ with the points ${\bf
A}_{1},{\bf A}_{3},{\bf A}_{5}$ (and also between  ${\bf
A}_{4},{\bf A}_{2},{\bf A}_{6}$)  has the same form. The constants
$\alpha_z=\Phi^{AB}_{zz}({\bf B}_{1}),\,
\gamma_z=\Phi^{AA}_{zz}({\bf A}_{1}),\,
\alpha_z^{\prime}=\Phi^{AB}_{zz}({\bf B}_{1}^{\prime}),\,
\alpha=\Phi^{AB}_{\xi\eta}({\bf B}_{1}),\,
\alpha^{\prime}=\Phi^{AB}_{\xi\eta}({\bf B}_{1}^{\prime}),$ and
$\gamma=\Phi^{AA}_{\xi\eta}({\bf A}_{1})$
are evidently real. The constant $\beta=\Phi^{AB}_{\xi\xi}({\bf
B}_{1})$ as well as $\beta^{\prime}=\Phi^{AB}_{\xi\xi}({\bf
B}_{1}^{\prime}))$ is real because the reflection $(x, y)
\rightarrow (x, -y)$ with ${\bf B}_{1}\rightarrow{\bf B}_{1}$,\,
${\bf B}_{1}^{\prime}\rightarrow{\bf B}_{1}^{\prime}$ belongs to
the symmetry group. Besides,  we have one complex force constant
$\delta=\Phi^{AA}_{\xi\xi}({\bf A}_{1})$.

Two  force constants $\Phi^{AA}_{zz}({\bf A}_{0})$
 and $\Phi^{AA}_{\xi\eta}({\bf
A}_{0})$ for the atom ${\bf A}_{0} $  can be excluded in the
ordinary way with the help of conditions imposed by invariance
with respect to the translations of the layer as a whole in the
$x/z$ directions. Using the equations of motion (\ref{eqmo}) and
Eqs. (\ref{sc}), (\ref{ss}), we find this stability condition
 $ \Phi^{AA}_{\xi
\eta}({\bf A}_{0})+6\Phi^{AA}_{\xi\eta}({\bf
A}_{1})+3\Phi^{AB}_{\xi \eta}({\bf B}_{1})+3\Phi^{AB}_{\xi
\eta}({\bf B}_{1}^{\prime})=0$
and  the similar form for the $zz$ components.

\subsection{ Dispersion of  bending out-of-plane modes} The out-of-plane
vibrations $u_z^{A}, u_z^{B}$ in the $z$ direction
 are not coupled with the in-plane modes because the force constants of  type
 $\Phi_{xz}$ or $\Phi_{yz}$ equals zero due to the reflection  $z\to -z$.
 The corresponding dynamical matrix
 for the out-of-plane modes has the form
\begin{equation}
\label{hamz} \left(\begin{array}{cc}
 \phi^{AA}_{zz}({\bf q}) &\phi^{AB}_{zz}({\bf q})\\
\phi^{AB}_{zz}({\bf q})^* &\phi^{AA}_{zz}({\bf q})
\end{array}\right ),
\end{equation}
where
 \begin{equation}\label{phz} \begin{array}{c}
\phi^{AA}_{zz}({\bf q})=-3(\alpha_z+\alpha_z^{\prime})\\
+2\gamma_z[\cos{(\sqrt3q_y)}+2\cos{(3q_x/2)}
\cos{(\sqrt3q_y/2)}-3]\,,\\
 \phi^{AB}_{zz}({\bf q})= \alpha_z
[\exp{(iq_x)}+2\exp{(-iq_x/2)} \cos{(\sqrt3q_y/2)}]\\
+\alpha_z^{\prime} [\exp{(-2iq_x)}+2\exp{(iq_x)}
\cos{(\sqrt3q_y)}]\,.
\end{array}
\end{equation}
The phonon dispersion for the out-of-plane
modes is found
\begin{equation} \label{pdz}
\omega_{\text{ZO,ZA}}({\bf q})=\sqrt{\phi^{AA}_{zz}({\bf q})\pm
|\phi^{AB}_{zz}({\bf q})|}.
\end{equation}
The equations allow  us to express the phonon frequencies of the
out-of-plane branches at the critical points $\Gamma , K$, and $M$
in terms of the force constants:
\begin{equation}
 \begin{array}{c}
 \omega_{\text{ZO}}(\Gamma)=\left[-6(\alpha_z+\alpha_z^{\prime})\right]^{1/2}
 \,\\ \nonumber
\omega_{\text{ZO, ZA}}(K)=\left[-3(\alpha_z+\alpha_z^{\prime})-9\gamma_z\right]^{1/2}\,\\
\omega_{\text{ZO}}(M)=\left[-4\alpha_z-8\gamma_z\right]^{1/2}\, \\
\omega_{\text{
ZA}}(M)=\left[-2\alpha_z-6\alpha_z^{\prime}-8\gamma_z\right]^{1/2}\,
.
\end{array}\label{gz}\end{equation}

 Expanding Eq. (\ref{pdz}) in powers of the wave vector $\bf{q}$, we find
  the velocity of the acoustic out-of-plane mode propagating in the
layer
\begin{equation}\label{sz}
s_{z}=a\left[-0.75\alpha_z-3\alpha_z^{\prime}-
4.5\gamma_z\right]^{1/2}= \sqrt{C_{44}/\rho},
\end{equation}
where we use the well-known formula for the velocity of the
acoustic $z$-mode propagating in the $x$-direction in terms of the
elastic constant $C_{44}$ and  density $\rho$ of a hexagonal
crystal. Because the interaction between the layers in graphite is
weak, we can correspond the values of $C_{44}$ and   $\rho$  to
graphite.
\begin{table}[] \caption{\label{tb1} Force constants in
10$^{5}$  cm$^{-2}$ : $\alpha$, $\beta$, and $\alpha_z$ for the
first neighbors\,; $\gamma$, $\delta$,  and $\gamma_z$ for the
second neighbors\,; $\alpha^{\prime}$, $\alpha^{\prime}_z$, and
$\beta^{\prime}$ for the third neighbors.}
        \begin{ruledtabular}
                \begin{tabular}{|c|c|c|c|c|c|c|c|c|}
     $\alpha$ &$\beta$&$\gamma$ &$\delta$&$\alpha^{\prime}$&$\beta^{\prime}$   &$\alpha_z$ &$\gamma_z$&$\alpha_z^{\prime}$\\
\hline
 -4.095& -1.645&-0.209&0.690&-0.072&0.375&-1.415&0.171&0.085
\end{tabular}
\end{ruledtabular}
\end{table}
 \subsection{ Dispersion of  in-plane modes}
The dynamical matrix for the in-plane vibrations has the form
similar to that for the in-plane mode (\ref{hamz}), but instead of
the functions $\phi_{zz}^{AA}({\bf q})$ and $\phi_{zz}^{AB}({\bf
q})$ we have to substitute correspondingly the $2\times2$ matrices
\begin{eqnarray}\label{hamm}
 \left ( \begin{array}{cc}
 \phi^{AA}_{\xi\eta}({\bf q})&\phi^{AA}_{\xi\xi}({\bf q})\\
   \phi^{AA}_{\xi\xi}({\bf q})^*&\phi^{AA}_{\xi\eta}({\bf q})
\end{array}\right ),\,
\left ( \begin{array}{cc}
  \phi^{AB}_{\xi\eta}({\bf q})&\phi^{AB}_{\xi\xi}({\bf q})\\
   \phi^{AB}_{\eta\eta}({\bf q})&\phi^{AB}_{\xi\eta}({\bf q})
\end{array}\right ).
\end{eqnarray}
The  matrix elements $\phi^{AA}_{\xi\eta}({\bf q})$ and
$\phi^{AB}_{\xi\eta}({\bf q})$ are obtained from
$\phi^{AA}_{zz}({\bf q})$ and $\phi^{AB}_{zz}({\bf q})$, Eqs.
(\ref{phz}), correspondingly, with substitutions  $\gamma$,
$\alpha$, and $\alpha^{\prime}$ instead of $\gamma_z$, $\alpha_z$,
and $\alpha_z^{\prime}$. The off-diagonal elements are given by
\begin{eqnarray}\nonumber
&\phi^{AA}_{\xi\xi}({\bf q})=\\\nonumber &\delta
[\exp(i\sqrt{3}q_y)+2\cos{(3q_x/2+2\pi/3)}\exp(-i\sqrt3q_y/2)]+\\
 \nonumber &\delta^*
[\exp(-i\sqrt{3}q_y)+2\cos{(3q_x/2-2\pi/3)}\exp{(i\sqrt3q_y/2)}],\\
\nonumber &\phi^{AB}_{\xi\xi}({\bf q})= \\\nonumber &\beta
[\exp{(iq_x)}+2\exp{(-iq_x/2)}
\cos{(\sqrt3q_y/2-2\pi/3)}]\\\nonumber &+ \beta^{\prime}
[\exp{(-2iq_x)}+2\exp{(iq_x)} \cos{(\sqrt3q_y+2\pi/3)}].
\end{eqnarray}
The matrix elements for the $B$ sublattice  can be obtained from
 that  for the $A$ sublattice  by  $C_2$ rotation
$(x,y)\rightarrow -(x,y)$ of the graphene symmetry group.
\begin{table}[]
\caption{\label{tb} Elastic constants (in 10 GPa) and the sound
velocities (in km/s) calculated (theo) and observed  (exp) . }
        \begin{ruledtabular}
                \begin{tabular}{|c|c|c|c|c|c|c|}
      &$C_{11}$ &$C_{12}$ &$C_{44}$&$s_{\text{LA}}$ &$s_{\text{TA}}$ &$s_z$\\
\hline
 theo&$ 86$ &$18$ &$0.57$&$ 19.5$ &$12.2$ &$1.59$\\
\hline
exp &$106\pm2^a$ &$18\pm2^a$ &$0.45\pm.05^a $&$\approx24^b$ &$14^b$ &\\
\end{tabular}
\end{ruledtabular}

$^a$ Reference \cite{GP},\,$^b$ Reference \cite{SAS,OAS},
\end{table}
 The optical phonon frequencies for the
in-plane branches at $\Gamma$ and $K$  are found
\begin{eqnarray}\label{pfx}\nonumber
\omega_{1,2}^{in-pl}(\Gamma)=[-6(\alpha+\alpha^{\prime})]^{1/2},\quad
\text{doublet},\\
\omega^{in-pl}_{1,2}(K)=[-3(\alpha+\alpha^{\prime})-9\gamma]^{1/2},\quad \text{doublet},\\
\nonumber
\omega^{in-pl}_{3,4}(K)=[-3(\alpha+\alpha^{\prime})-9\gamma\pm
3(\beta+\beta^{\prime})]^{1/2}\,.
\end{eqnarray}
 An algebraic
equation of the forth order have to be solved in order to find the
phonon frequencies at the M point as well as at points of the
general position.
\begin{figure}[b]
\resizebox{.5\textwidth}{!}{\includegraphics{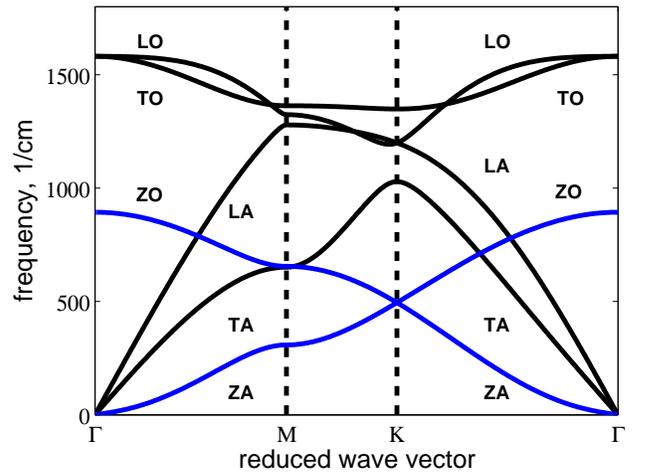}}
\caption{ \label{2} Calculated phonon dispersion for graphene; the
force constants, elastic constants, and  phonon frequencies at
critical points are listed   in Tables 1, 2, and 3
correspondingly. }
\end{figure}
\begin{table*}[]
\caption{\label{tb3} Phonon frequencies at critical points in
cm$^{-1}$; $z$ and $\parallel$ stand for the out-of-plane and
in-plane branches, respectively. }
        \begin{ruledtabular}
                \begin{tabular}{|c|c|c|c|}
   &$\Gamma$ \, [0 0]    &$M$  \,   [1 $\sqrt3]\pi/3a$    &$K$\, [0 1]$4\pi/3\sqrt3 a$ \\
 \hline
&$\omega^{\parallel}$\qquad$\omega^{z}$&$\omega^{\parallel}_1$\qquad
$\omega^{\parallel}_2$\qquad$\omega^{\parallel}_3$\qquad$\omega^{\parallel}_4$\qquad$\omega^{z}_1$
\quad
$\omega^{z}_2$&$\omega^{\parallel}_1$\qquad $\omega^{\parallel}_{2,3}$\qquad $\omega^{\parallel}_4$\quad$\omega^{z}_{1,2}$\\
\hline
   exp&1590$^a$\quad861$^a$\quad&\,1389$^a$\qquad\qquad\quad
\quad\quad
630$^d$\quad 670$^a$\quad 471$^c$ &1313$^d$\, 1184$^b$\quad \quad\quad 482$^d$\quad \\
      &1583$^b$\quad 868$^c$\quad&1390$^b $\, \,1323$^b \,$\,
1290$^b\,\,$\,\,
\qquad\quad \qquad\quad 451$^d$ &1265$^b$\, 1194$^b$\quad \quad\quad 517$^d$\quad \\
      &1565$^b$\quad 868$^e$\quad&\qquad\qquad \qquad \qquad\quad
\quad625$^e$\quad625$^e$\quad 480$^e$ & 1285$^e$ \qquad\quad 1021$^e$\, 537$^e$ \\
\hline
             theo$^{f}$&1595 \quad890\quad&1442\quad 1380\quad\,1339\quad\,
\,636\,\quad618\qquad 475&1371\quad 1246\quad \,994\,\quad 535\quad \\
             theo$^{b}$&\, 1581 \quad\qquad\qquad&1425\quad 1350\quad
\,1315\quad\,
\qquad\quad\quad\qquad\qquad\qquad& 1300\quad 1220\,\quad\qquad\quad\quad\quad\\
             theo&1581 \quad893\quad&1363\quad 1324\quad \,1279\quad\,
\,651\,\quad655\qquad 308&1349\quad 1199\quad 1028\quad 495\quad \\
\end{tabular}
\end{ruledtabular}
 $^a$ Reference \cite{OA}, $^b$ Reference \cite{MRTR}, $^c$
Reference \cite{NWS}, $^d$ Reference \cite{Y}, $^e$ Reference
\cite{MMD}, $^f$ Reference \cite{DK}
\end{table*}
{

The in-plane vibrations make a contribution into the elastic
constants C$_{11}$ and C$_{12}$. The corresponding relation
between the dynamic matrix elements and the elastic constants  can
be deduced taking the long-wavelength limit (${\bf q}\rightarrow
0$) in the matrices (\ref{hamm}). In this limit, separating the
acoustic vibrations ${\bf u}^{\text{ac}}$ from the optical modes,
we obtain the equation of motion in the matrix form
\begin{equation}\label{ac}\begin{array}{c}
\left[(\phi ^{AA}+\phi ^{AB}+ \phi ^{BB}+ \phi ^{BA})/2\right.\\+
\left.\phi _1^{AB} (\phi _0^{AB})^{-1} \phi _1^{AB}
-\omega^2\right] {\bf u}^{\text{ac}}=0,
\end{array}\end{equation}
where the subscripts 0 and 1 mean that  the terms of the zero and
first order in ${\bf q}$ should, correspondingly,  be kept in the
matrices (\ref{hamm}), but the expansion to the second order is
used in other terms. We  find the matrix factor of ${\bf
u}^{\text{ac}}$ in Eqs. (\ref{ac}):
\begin{eqnarray}\label{acm}
 \left ( \begin{array}{cc} s_1q^2-\omega^2& s_2q_{+}^2\\
                                       s_2q_{-}^2 & s_1q^2-\omega^2
\end{array}\right ),\end{eqnarray}
where
\begin{eqnarray}\nonumber
&s_1=-\frac{9}{2}\gamma-\frac{3}{4}\alpha-3\alpha^{\prime}+\frac{3}{8}
(\beta-2\beta^{\prime})^2/(\alpha+\alpha^{\prime}),\\\nonumber
 &s_2=\frac{9}{4}\text{Re}( \delta)-\frac{3}{8}\beta-\frac{3}{2}\beta^{\prime}.
  \end{eqnarray}
With the help of Eq. (\ref{acm}),  we obtain the  velocities of
longitudinal and transverse acoustic in-plane modes
\begin{equation}\label{lv}
 \begin{array}{c}
 s_{\text {LA}}=a\sqrt{s_1+s_2}=\sqrt{C_{11}/\rho},\,\\
 s_{\text {TA}}=a\sqrt{s_1-s_2}=\sqrt{(C_{11}-C_{12})/2\rho},
\end{array}\end{equation}
corresponding them to the  elastic
 constants $C_{11},\,C_{12}$ and  density $\rho$ of graphite.

\section{Results and discussions }

The calculated phonon dispersion  is shown in Fig. 2. Notice,
first, that the sound velocities (for  long waves,
$q\rightarrow\Gamma$) are isotropic in the $xy$ plane as it should
be appropriate for the  symmetry of graphene. Second, the in-plane
LO/TO modes at $\Gamma$, the in-plane LO/LA modes at $K$, and the
out-of-plane ZA/ZO modes at $K$ are doubly degenerate because
graphene is the non-polar crystal and the $C_{3v}$ symmetry of
these points in the Brillouin zone admits the two-fold
representation (observation of splitting of that modes in graphene
would  display the symmetry braking of the crystal).

Because of the lack of information on graphene, we  compare the
present theory with experiments on graphite. Thus, we have only
three force constants $\alpha_z$, $\gamma_z$, and
$\alpha^{\prime}_z$ to fit four  frequencies of the out-of-plane
modes at the critical points $\Gamma$, $M$, and $K$. We must keep
in mind that the frequencies in graphene for the out-of-plane
branches could be less than their values in graphite since the
atoms are more free to move in the $z$ direction in graphene
comparatively with graphite. It is evident that  the adjacent
layers in graphite affect  the low frequencies more intensively.
The interaction of the adjacent layers can be estimated   from the
 ZA -- ZO$^{\prime}$  splitting about 130 cm$^{-1}$ given,
 for instance, in Ref. \cite{MM}. These modes become degenerate
  when the inter-layer interaction is switched off.
  Therefore, the  lowest
frequencies of out-of-plane modes calculated at the $M$ and $K$
points are considerably less than the corresponding frequencies
observed in graphite (see Table 3).

Furthermore, the force constants determine the velocity $s_z$, Eq.
(\ref{sz}), of the acoustic out-of-plane mode along with the
elastic constant $C_{44}$. We  see that the velocity has the
nonzero value  unless a definite condition is satisfied for the
force constants. Using the values of force constants obtained in
fitting with the experimental data (see Table 1), we find the
value of the sound velocity $s_z=1.59$ km/s for the out-of-plane
mode. This  result is contradictory to the  statement of Ref.
\cite{DS} that the acoustic out-of-plane mode has a quadratic
dispersion. The fact that the sound velocity $s_z$ is very
sensitive to the small variation of $\gamma_z$  indicates that
graphene is nearly unstable with respect to transformation into a
phase of the lower symmetry group at $\Gamma$.

For the in-plane modes, we have to fit eight  frequencies at the
critical points and two elastic constants.  Equations (\ref{pfx})
and (\ref{lv}) can be used as a starting point.
 Fitting of the in-plane branches is insensitive to the imaginary
part of the constant  $\delta$. Therefore, it is taken as a real
parameter. Results of the fit are presented in Fig. 2 and Tables.
 Notice, that the
extent of agreement of the present theory with the  data obtained
for graphite corresponds to the comparison level between the
first-principle calculations  for graphite  in Ref. \cite{MRTR}
and their experimental data (see Table 3). The largest
disagreement of 5\% between  our calculations and  experiments on
graphite for the highest phonon mode  occurs at the K point. This
is result of the Kohn anomaly   due to  the electron--phonon
interaction \cite{PLM} which  reduces the  phonon frequency at K.
 The same reason explains some overbending observed probably in
graphite along the $\Gamma -M$ direction.
\section {Conclusions} We calculate the phonon dispersion in
graphene using the Born--von Karman model with the first-,
second-, and  third-neighbor interactions imposed by the symmetry
constraints. The bending (out-of-plane) modes are not coupled with
the in-plane branches and indicate the latent instability of
graphene with respect to transformation into a lower-symmetry
phase. The acoustic ZA mode has the linear dispersion in a small
wave-vector interval near the $\Gamma$ point. The optical
frequencies of these modes are less than the corresponding values
in graphite. For the higher in-plane modes, the fit shows good
agreement between the experimental and calculated values of
optical frequencies,   elastic constants, and acoustic velocities.

\acknowledgements
 The work was supported by the Russian Foundation
for Basic Research  (grant No.07-02-00571).


\begin{references}
\bibitem{Novo} K.S. Novoselov, A.K. Geim, S.V. Morozov et al.,
Science, \textbf{306, }666 (2004); K.S. Novoselov  et al., Nature,
\textbf{438, }197 (2005).
\bibitem{ZSA} Y. Zhang, J.P. Small, M.E.S. Amory, and P.Kim,
Phys. Rev. Lett. \textbf{94}, 176803 (2005).
\bibitem{FMS} C.C. Ferari,
 J.C. Meyer, V. Scardaci, C. Caseraghi, M. Lazzeri, F. Mauri, S. Piscanec,
D. Jiang, K.S. Novoselov, S. Roth, and A.K. Geim, Phys. Rev. Lett.
\textbf{97}, 187401 (2006).
\bibitem{NG} A.H. Castro Neto, F. Guinea,
Phys. Rev. B \textbf{75}, 045404 (2007).
\bibitem{DL} J. De Launay, Solid State Phys. \textbf{3}, 203 (1957).
\bibitem{NWS} R. Nicklow, W. Wakabayashi, and H.G. Smith,
Phys. Rev. B \textbf{5}, 4951 (1972).
\bibitem{AR} A.A. Ahmadieh and H.A. Rafizadeh,
Phys. Rev. B \textbf{7}, 4527 (1973).
\bibitem{NB} A.P.P. Nicholson and D.J. Bacon, J. Phys. C
\textbf{10}, 2295 (1977).
\bibitem{MKH} M. Maeda, Y. Kuramoto, and C. Horie, J. Phys. Soc. Jpn. Lett.
 \textbf{47}, 337 (1979).
\bibitem{AD} R. Al-Jishi and G. Dresselhaus,
Phys. Rev. B \textbf{26}, 4514 (1982).
\bibitem{GMR} H. Gupta, J. Malhotra, N. Rani, and B. Tripathi,
Phys. Rev. B \textbf{33}, 7285 (1986).
\bibitem{MMD} M. Mohr,
J. Maultzsch, E. Dobard\u{z}i\'{c}, I. Milo\u{s}evi\'{c}, M.
Damnjanovi\'{c}, A. Bosak, M. Krish, and C. Thomsen, 
Phys. Rev. B \textbf{76}, 035439 (2007).
\bibitem{LDC} L. Lang, S. Doyen-Lang, A. Charlier, and M.F.
Charlier,  Phys. Rev. B \textbf{49}, 5672 (1994).
\bibitem{BO} G. Benedek and G. Onida,  Phys. Rev. B \textbf{47}, 16471 (1992).
\bibitem{MCZ} C. Mapelli, C. Castiglioni,  G. Zerbi, and K. M\"{u}llen,
Phys. Rev. B \textbf{60}, 12710 (1999).
\bibitem{OA}  T. Aizava, R. Souda, S.Otani, Y. Ishizava, and C. Oshima,Y.
Samiyosh, Phys. Rev. B \textbf{42}, 11469 (1990).
\bibitem{DK} O. Dubay and G. Kresse, Phys. Rev. B \textbf{67}, 035401 (2003).
\bibitem{WR} L. Wirtz and A. Rubio, Solid State Commun. \textbf{131}, 141 (2004).
\bibitem{MRTR} J. Maultzsch,
S. Reich, C. Thomsen, H. Reequardt, and P. Ordejon, Phys. Rev.
Lett. \textbf{92}, 075501 (2004).
\bibitem{MM} N. Mounet and N. Marzari,  Phys. Rev. B \textbf{71}, 205214 (2005).
\bibitem{PL} V.N. Popov and P. Lambin,
 Phys. Rev. B \textbf{73}, 085407 (2006).
\bibitem{DS} R. Saito, G. Dresselhaus, and M.S. Dresselhaus,  {\it  Physical Properties of Carbon
 Nanotubes,}  p. 170, (Imperial College Press, London, 2003).
\bibitem{GP} {\it Graphite and Precursors}, ed. by P. Delhaes
(Gordon and Breach, Australia, 2001), Chap. 6.
\bibitem{SAS} D. S\'{a}nchez-Portal, E. Artacho,  J.M. Soler, A.Rubio, and P. Ordej\'{o}n, Phys. Rev. B \textbf{59}, 12678 (1999).
\bibitem{OAS} C. Oshima, T. Aizava, R. Souda, Y. Ishizava, and Y.
Samiyosh, Solid. State. Commun. \textbf{65}, 1601 (1988).
\bibitem{Y} H. Yanagisawa,
T. Tanaka, Y. Ishida, M. Matsue, E. Rokuta, S. Otani, and C.
Oshima, Surf. Interface Anal. \textbf{37}, 133 (2005).
\bibitem{PLM} S. Piscanec, M. Lazzeri, F. Mauri, A.C. Ferrari,
and J. Robertson, Phys. Rev. Lett. \textbf{93}, 185503 (2004).

\end{references}
\end{document}